\documentclass[prl,twocolumn,showpacs]{revtex4}
\usepackage{graphicx}
\usepackage{bm}
\usepackage{amssymb,amsfonts,amsmath}

\newcommand{\refeqn}[1]{(\ref{#1})}
\newcommand{\reffig}[1]{FIG. \ref{#1}}

\renewcommand{\d}{\mathrm{d}}
\newcommand{\e}{\mathrm{e}}
\renewcommand{\Re}{\mathrm{Re}\;}

\begin{document}
\title{Collective oscillation in two-dimensional fluid}
\author{Hidetoshi Morita}
\email{hidetoshi.morita@scphys.kyoto-u.ac.jp}
\affiliation{Department of Physics, Kyoto University, 606-8502 Kyoto, Japan}

\date{\today}

\begin{abstract}
Large-scale collective oscillation
is discovered in the two-dimensional Euler equations.
For initial conditions far from a base stationary flow,
the system does not relax to the base stationary flow,
but instead shows pairs of coherent vortices moving along the base stream line,
which leads to large-scale oscillatory fields.
The investigation of the vicinity of a bifurcation point
suggests that this oscillation appears through Hopf bifurcation.
Furthermore, a dynamic self-consistent theory explains
that this oscillation results from the collective organization
of a state of self-oscillation.
\end{abstract}


\maketitle

Two-dimensional (2D), and similarly geophysical, flows of fluids
with high Reynolds numbers are widely observed in nature.
Examples include
oceanic and atmospheric flows~\cite{Pedlosky}.
2D flows are also matters of experiment~\cite{Sommeria_1986}.
A characteristic phenomenon in such flows is large-scale pattern formation.
Jupiter's red spot and Kuroshio of the north pacific ocean
are well-known examples in nature.
Although highly turbulent on intermediate and small scales,
stationary patterns of flow appear on large scales.
This is due to the so-called inverse cascade~\cite{Kraichnan_1967},
which is an essential difference from three-dimensional turbulence.

A theoretical interest lies in
the statistical mechanics description of the large-scale patterns
\cite{Miller_1990,Robert_1991},
which has been studied intensively over the last two decades
\cite{Majda_Wang}.
Instead of the infinite resolution description of the bare vorticity field,
a probabilistic description of the field on a coarse-grained space is introduced.
That is, the probability density function of the vorticity level
is defined at each point.
Then maximizing the corresponding Shannon entropy functional
yields the large-scale stationary pattern of flow.
With the use of this theory,
Jupiter's red spot, for example,
has been successfully explained~\cite{Bouchet_Sommeria_2002}.

Here we recall the general scheme of macroscopic systems.
Needless to say, equilibrium states are described
with statistical mechanics as entropy maximum states.
When the systems are non-equilibrium but close to equilibrium,
they show simple relaxation to, or fluctuation around, the equilibrium states.
When the systems are far from equilibrium,
they often show temporal motions on macroscopic scales,
such as periodic oscillation,
by spontaneously breaking the temporal symmetry~\cite{Prigogine}.

In this context, we reflect the 2D fluids.
The large-scale \textit{stationary} patterns are regarded as
the \textit{equilibrium} states of statistical mechanics, as mentioned above.
When the systems are slightly far from the stationary flows,
they show relaxation to
\cite{Robert_Sommeria_1992,Bouchet_Morita_2010},
or fluctuation around
\cite{Bouchet_Simonnet_2009,Morita_Simonnet_Bouchet},
the stationary flows,
though their behaviors are not so simple as ordinary thermodynamic systems.
Then, by thinking parallelly to the above route to non-equilibrium,
we expect large-scale temporal motions also in the 2D fluids.
Indeed, it has been reported that macroscopic oscillation appears,
through Hopf bifurcation, in the 2D Navier-Stokes equations
under external shear force~\cite{Okamoto_Shoji_1993}.
However, this shear obviously breaks the symmetry,
which may cause temporal motions.
Then, as a nontrivial question,
it is natural to ask if the large-scale non-stationary motions exist
without net external forcing.
The present paper reports the discovery of
such large-scale temporal oscillation in the 2D Euler equations.

We consider the 2D Euler equations,
\begin{align}
\partial_t\omega+\bm{v}\cdot\nabla\omega=0,
\label{eqn:Euler}
\end{align}
on a doubly periodic domain
$\mathcal{D}=[-L_x/2,L_x/2)\times[-L_y/2,L_y/2)$.
Here $\bm{v}=(v_x,v_y)$ is the velocity field,
and $\omega=\partial_x v_y-\partial_y v_x$ is the vorticity field.
We set $L_x=2\pi$ and $L_y=2\pi\Gamma$,
with the aspect ratio $\Gamma >1$,
without loss of generality.

The large-scale motion is realized
when the initial condition is set far from the stationary state
obtained from the statistical mechanics description.
We give initial conditions in the form,
\begin{align}
\omega(0,\bm{r})=\Omega(\bm{r})+\delta\omega(0,\bm{r}),
\label{eqn:w_W_dw}
\end{align}
where $\Omega(\bm{r})$ is a stationary base flow
and $\delta\omega(0,\bm{r})$ is an initial perturbation.

As the base flow, we consider the so-called zonal flow,
or the Kolmogorov flow,
of the form,
\begin{align}
\Omega(\bm{r})=-\frac{2\pi}{L_y}\cos{\frac{2\pi}{L_y}y}
\label{eqn:W_def}
\end{align}
It is easily shown that this is a stationary solution of the 2D Euler equations.
Moreover, this is the entropy maximum state
in the statistical mechanics description
of the 2D Euler equations on a doubly periodic domain
\cite{Yin_Montgomery_Clercx_2003,Bouchet_Simonnet_2009}.

Since we are interested in large-scale behaviors,
it is natural to apply the initial perturbations on the largest scale.
We consider the initial perturbations of the form,
\begin{align}
\delta\omega(t=0,\bm{r})=-\epsilon \frac{2\pi}{L_x}\cos{\frac{2\pi}{L_x}x}
\label{eqn:dw_def}
\end{align}
where $\epsilon$ represents
the distance of the initial condition from the stationary state,
which is not necessarily small.

In the numerical simulations,
we add a very weak hyper-viscous term $(-1)^{h+1}\nu\nabla^{2h}\omega$
to the r.h.s. of \refeqn{eqn:Euler},
with $h=4$ and $\nu=2\cdot 10^{-18}$,
just to stabilize the numerical scheme.
We use the pseudo-spectral method \cite{Orszag},
which is known as the most precise and robust numerical algorithm,
and is indeed regarded as the standard method,
for the 2D Euler equations in doubly periodic domains.
We use the resolution $256\times 256$.

After a short transient ($t \le 200$),
the system goes into stable states.
We find three sorts of states,
depending on the parameters ($\epsilon$, $\Gamma$).
To quantatively distinguish these states,
we introduce an order parameter
that is the Fourier component of the smallest wavenumber,
\begin{align}
Z(t)=-\Re\hat{\omega}_{(1,0)}(t)
\end{align}
where
$
\hat{\omega}_{\bm{q}}(t)
=\int_\mathcal{D}\frac{\d^2\bm{r}}{|\mathcal{D}|}\;\omega(t,\bm{r})\e^{-i\bm{q}\cdot\bm{r}}.
$
Note that the imaginary part
is kept zero for the present initial conditions.

\begin{figure}[htb]
\includegraphics[width=0.47\textwidth]{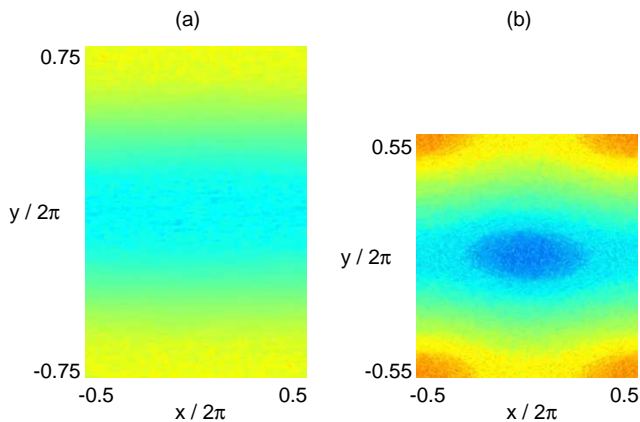}
\caption{
Snapshots of the relaxed vorticity field;
(a) zonal ($\Gamma=1.5$, $\epsilon=0.1$)
and (b) dipole ($\Gamma=1.1$, $\epsilon=0.5$) flow.
}
\label{fig:w_st}
\end{figure}

\begin{figure}[htb]
\includegraphics[width=0.45\textwidth]{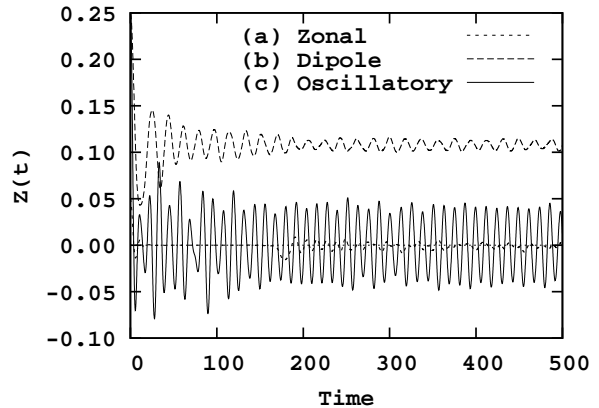}
\caption{
Typical time series of the order parameter $Z(t)$,
for (a) $(\Gamma,\epsilon)=(1.5,0.1)$ (zonal),
(b) $(1.1,0.5)$ (dipole), and (c) $(1.5,0.5)$ (oscillatory).
}
\label{fig:w10_tseri}
\end{figure}

One of the states is stationary zonal flow (\reffig{fig:w_st}-(a)),
similarly to the base flow on the largest scales.
Accordingly, $Z(t)$ is almost zero,
with small fluctuations (\reffig{fig:w10_tseri}-(a)).
Another state is stationary dipole flow (\reffig{fig:w_st}-(b));
the vorticity field shows a dipole pattern on the largest scales.
Accordingly, $Z(t)$ is almost kept to a nonzero value,
with small fluctuations (\reffig{fig:w10_tseri}-(b)).
Note that these two patterns are relevant
to the entropy maximum states in the statistical mechanics description
\cite{Yin_Montgomery_Clercx_2003,Bouchet_Simonnet_2009}.
These mean that, on the largest scales,
the system relaxes (strictly, close) to the entropy maximum state,
just as an ordinary thermodynamic system relaxes to equilibrium.

\begin{figure}[htb]
\includegraphics[width=0.45\textwidth]{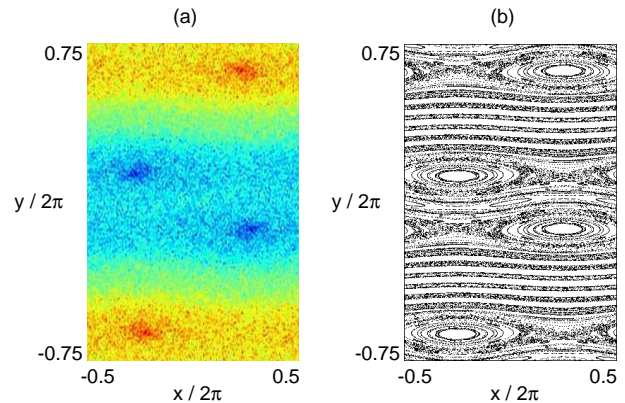}
\caption{
(a) A snapshot of the vorticity field in the 2D Euler equations
$(\Gamma=1.5, \epsilon=0.5)$.
and
(b) the Poincar\'e section in the one-body dynamics
\refeqn{eqn:eqn_of_mot_per} ($b=0.042$, $T=12$),
when the phase of oscillation is zero.
$\Gamma=1.5$.
}
\label{fig:w_compare}
\end{figure}

The other state, and what we are interested in,
is the non-stationary pairwise motion of coherent vortices
(\reffig{fig:w_compare}-(a)).
Two pairs of positive and negative coherent vortices
point-symmetrically move along the stream line of the base flow.
The motion of these vortices leads to oscillatory field.
Indeed, $Z(t)$ shows oscillation with a rather large amplitude
(\reffig{fig:w10_tseri}-(b)).
This is in a strong contrast to the above two states
in that no relevant state exists
in the statistical mechanics description.

\begin{figure}[htb]
\includegraphics[width=0.25\textwidth]{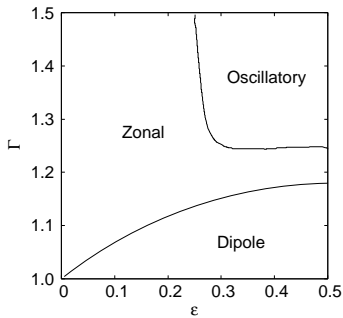}
\caption{
The phase diagram.
}
\label{fig:ph_diag}
\end{figure}

By changing the parameters $\epsilon$ and $\Gamma$,
we obtain the phase diagram, which is shown in \reffig{fig:ph_diag}.

\begin{figure}[htb]
\includegraphics[width=0.4\textwidth]{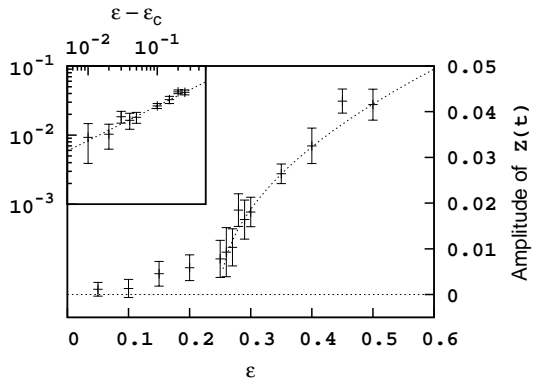}
\caption{
The bifurcation diagram.
The amplitude of oscillation of $Z(t)$
against $\epsilon$ and $\epsilon-\epsilon_c$ (inset in log-log plot),
with the curve proportional to $(\epsilon-\epsilon_c)^{1/2}$,
where we obtain $\epsilon_c=0.25$. $\Gamma=1.5$.
The error bars denote standard deviations of the small fluctuations.
}
\label{fig:bif_diag}
\end{figure}

The system shows a transition
from the stationary zonal flow to the oscillatory flow.
Then we investigate how this transition occurs.
\reffig{fig:bif_diag} shows the amplitude of oscillation of $Z(t)$
against the intensity of perturbation $\epsilon$.
The amplitude shows a continuous bifurcation at a critical point.
In particular, the amplitude increases proportionally to
$(\epsilon-\epsilon_c)^{1/2}$, where $\epsilon_c$ is the critical point.
This result suggests that the bifurcation from the zonal to oscillatory flow
is a Hopf-type bifurcation.

Next,
we investigate the mechanism how the large-scale oscillation is kept stable,
by a dynamic self-consistent theory
\cite{Morita_Kaneko_2006}.
Suppose
an ensemble of independent ``test point vortices''
moving along the stream line of
a temporary oscillating auxiliary field on the largest scale.
The collective motion of the point vortices results in a dynamic field
in the limit of infinite number of the vortices.
If these two fields,
one being the cause and the other the effect,
are identical,
this system is no different
from the original autonomous 2D Euler equations.

Suppose, on $\mathcal{D}$,
the temporally periodic auxiliary vorticity field
on the largest scale in the form,
\begin{align}
\omega^{\mathrm{aux}}(t,\bm{r})=
-\Omega(\bm{r})
-2Z^{\mathrm{aux}}(t)
\cos\frac{2\pi x}{L_x}
\end{align}
where $Z^{\mathrm{aux}}(t)$ is the order parameter of the auxiliary field
being sinusoidal with respect to time with the amplitude $b$ and period $T$,
$Z^{\mathrm{aux}}(t)=b\sin{2\pi t / T}.$
Each of the test point vortex
moves along its stream line, following the equations of motion,
\begin{subequations}
\begin{align}
\dot{x}=u_x^{\mathrm{aux}} &= +\partial_y\psi^{\mathrm{aux}} = \sin{\frac{2\pi y}{L_y}}
\\
\dot{y}=u_y^{\mathrm{aux}} &= -\partial_x\psi^{\mathrm{aux}} = -2Z^{\mathrm{aux}}(t)\frac{L_x}{2\pi}\sin{\frac{2\pi x}{L_x}}
\end{align}
\label{eqn:eqn_of_mot_per}
\end{subequations}
where $\psi_{aux}$ is the corresponding auxiliary stream function,
$\omega^{\mathrm{aux}}=-\nabla^2\psi^{\mathrm{aux}}.$

We consider this one-body dynamical system
with $b$ and $T$ the same value as $Z(t)$
numerically obtained from the original 2D Euler equations.
\reffig{fig:w_compare}-(b) shows the Poincar\'e plot of the trajectory
on $2\pi t/T\equiv 0\;(\!\!\!\!\mod2\pi)$,
which is similar to the standard map~\cite{Chirikov_1979}.
In between the region of fixed point ($y=0, L_y/2$)
and Kolmogorov-Arnold-Moser tori ($y=\pm L_y/4$),
there are the so-called chaotic seas.
In addition, within the chaotic seas, we find
two pairs of large 1-to-1 resonance island of tori.

We compare in \reffig{fig:w_compare}
this Poincar\'{e} section of the one-body dynamics (b)
with the snapshot of the vorticity field of the original 2D Euler equations (a)
when the phase of oscillation is each zero.
We notably find that the position of the coherent vortices
in the 2D Euler equations
coincides to that of 
the 1:1 resonance island of tori
in the one-body dynamics;
$(x/L_x,y/L_y)=(\pm 0.25,\mp 0.125), (\pm 0.25,\pm 0.625)$
at $2\pi t/T\equiv 0\;(\!\!\!\!\mod2\pi)$.

This coincidence self-consistently explains
the mechanism of the large-scale oscillation.
Once a considerable amount of vorticity is
concentrated on the 1:1 resonance islands of tori,
it leads to large-scale oscillatory field.
Thus generated oscillatory field, in turn, holds the vorticity
inside the islands,
and prevents it from diffusing away.
Thereby the state of self-oscillation is organized,
and the large-scale oscillation is stabilized.

We derive the above dynamic self-consistent solution on the largest scale.
Let $g(\bm{r})$
be a function rapidly decreasing and periodic on $\mathcal{D}$,
which express the distribution of the point vortices.
Since the concentrated vortices remain inside the 1-to-1 resonance islands of tori,
the distribution function is supposed to be advected along the stream line,
without changing its form, $g$, in the first approximation.
Then the vorticity field attributed to the point vortices is,
\begin{align}
\omega^{\mathrm{p.v.}}(t,\bm{r})-\Omega(\bm{r})=
-g(\bm{r}-\bm{R}_-)
-g(\bm{r}+\bm{R}_-)
\nonumber\\
+g(\bm{r}-\bm{R}_+)
+g(\bm{r}+\bm{R}_+)
\end{align}
where 
$\pm\bm{R}_\pm$ are the coordinates of the four resonance islands;
$\bm{R}_-=(-X,Y)$ and $\bm{R}_+=(-X+\frac{L_x}{2},-Y+\frac{L_y}{2})$.
In the first approximation,
$(X(t), Y(t))$ moves along the base stream line;
$Y(t)$ is constant with respect to time, $Y_0$,
and accordingly $X(t)$ changes in a constant speed,
$X(t)=-\frac{L_x}{4}+t\sin(2\pi Y_0/L_y)$.
Then the Fourier components of the vorticity field are,
\begin{align}
\hat{\omega}^{\mathrm{p.v.}}_{\bm{q}}(t)=&
\hat{\Omega}_{\bm{q}}+
\hat{g}_{\bm{q}}\{
-2\cos(q_x X-q_y Y)
\nonumber\\
&+2(-1)^{\frac{q_xL_x}{2\pi}}(-1)^{\frac{q_yL_y}{2\pi}}\cos(q_x X+q_y Y)
\}.
\end{align}
In particular,
$Z^{\mathrm{p.v.}}(t)=-\Re\hat{\omega}^{\mathrm{p.v.}}_{(1,0)}=4\hat{g}_{(1,0)}\cos X(t)=4\hat{g}_{(1,0)}\sin(t\sin(2\pi Y_0/L_y))$.
Thus the self consistent solution,
$\omega^{\mathrm{aux}}=\omega^{\mathrm{p.v.}}$,
is yielded, on the largest scale, as,
\begin{align}
b=2\hat{g}_{(1,0)},
\qquad
\frac{2\pi}{T}=\sin\frac{2\pi Y_0}{L_y}.
\end{align}

In conclusion,
we have discovered the non-stationary flow in the 2D Euler equations.
The motion of positive and negative coherent vortices
leads to the large-scale oscillation.
In the vicinity of the bifurcation point,
the amplitude of oscillation of the order parameter $Z(t)$
increases proportionally to $(\epsilon-\epsilon_c)^{1/2}$,
suggesting Hopf bifurcation.
This implies a structures of low-dimensional dissipative dynamical system
in the high-dimensional conservative dynamics of the 2D Euler equations.
By comparing the dynamics of these coherent vortices
in the original 2D Euler equations
with the one-body dynamics of the test point vortex
under the auxiliary oscillatory fields,
we have explained self-consistently that this oscillation
result from the collective organization of the state of self-oscillation.

We have focused our consideration only on the largest scale.
In turbulence, in general,
the large-scale and small-scale dynamics are not simply separated.
Indeed, as shown in \reffig{fig:w10_tseri},
neither the oscillatory nor the stationary state
is completely periodic nor constant, respectively,
but is accompanied by small fluctuations;
accordingly, the bifurcation is not completely clear (\reffig{fig:bif_diag}).
These small fluctuations are likely to be attributed to the oscillations
within the single resonance islands
\cite{small_osc}
as well as the motions of higher-to-higher resonances.
Taking larger wavenumber or larger frequency modes of field into account
could reproduce the oscillation more precisely.
Nevertheless, it is remarkable that still the mechanism
of the present large-scale oscillation is essentially explained
only with the large-scale description.

The collective oscillation discussed in this paper
has the same mechanism as that
in many-body Hamiltonian systems~\cite{Morita_Kaneko_2006}.
That is, both these two systems have long-range interactions
\cite{Dauxois_RUffo_Cugliandolo@LesHouches2008},
which indeed organize the state of self-oscillation.
Thus this collective oscillation may be a universal phenomenon
in long-range interacting conservative systems,
whether the system is a many-body or continuous system.
The benefit of the 2D fluid is that it is experimentally
confirmed more easily than the Hamiltonian systems.
Our initial condition will be realized as follows;
we prepare the system with the aspect ratio unity,
in which case the field
(\ref{eqn:w_W_dw}, \ref{eqn:W_def}, \ref{eqn:dw_def}) is stable,
and then we suddenly change the aspect ratio into $\Gamma$.

Finally, we also observe a similar motion of coherent vortices
in the 2D stochastic Navier-Stokes (SNS) equations
with weak Rayleigh friction and viscosity.
Starting from arbitrary initial conditions,
after a transient, the system self-organizes the dynamic pattern.
Due to the stochasticity, this pattern does not last so long,
but disappear in a rather short time.
However, it re-appears soon.
The annihilation and creation of these pairwise vortices repeats continually.

The author would like to thank Marianne Corvellec,
Eric Simonnet, and Shinji Takesue for comments.
This work was supported by
the Grant-in-Aid for the Global COE Program
``The Next Generation of Physics, Spun from Universality and Emergence''
from MEXT, Japan.

\bibliographystyle{apsrev}

\begin{thebibliography}{25}
\expandafter\ifx\csname natexlab\endcsname\relax\def\natexlab#1{#1}\fi
\expandafter\ifx\csname bibnamefont\endcsname\relax
  \def\bibnamefont#1{#1}\fi
\expandafter\ifx\csname bibfnamefont\endcsname\relax
  \def\bibfnamefont#1{#1}\fi
\expandafter\ifx\csname citenamefont\endcsname\relax
  \def\citenamefont#1{#1}\fi
\expandafter\ifx\csname url\endcsname\relax
  \def\url#1{\texttt{#1}}\fi
\expandafter\ifx\csname urlprefix\endcsname\relax\def\urlprefix{URL }\fi
\providecommand{\bibinfo}[2]{#2}
\providecommand{\eprint}[2][]{\url{#2}}

\bibitem[{\citenamefont{{Pedlosky}}(1982)}]{Pedlosky}
\bibinfo{author}{\bibfnamefont{J.}~\bibnamefont{{Pedlosky}}},
  \emph{\bibinfo{title}{{Geophysical fluid dynamics}}}
  (\bibinfo{publisher}{Springer-Verlag}, \bibinfo{year}{1982}).

\bibitem[{\citenamefont{Sommeria}(1986)}]{Sommeria_1986}
\bibinfo{author}{\bibfnamefont{J.}~\bibnamefont{Sommeria}},
  \bibinfo{journal}{J. Fluid Mech.} \textbf{\bibinfo{volume}{170}},
  \bibinfo{pages}{139} (\bibinfo{year}{1986}).

\bibitem[{\citenamefont{Kraichnan}(1967)}]{Kraichnan_1967}
\bibinfo{author}{\bibfnamefont{R.~H.} \bibnamefont{Kraichnan}},
  \bibinfo{journal}{Phys. Fluids} \textbf{\bibinfo{volume}{10}},
  \bibinfo{pages}{1417} (\bibinfo{year}{1967}).

\bibitem[{\citenamefont{Miller}(1990)}]{Miller_1990}
\bibinfo{author}{\bibfnamefont{J.}~\bibnamefont{Miller}},
  \bibinfo{journal}{Phys. Rev. Lett.} \textbf{\bibinfo{volume}{65}},
  \bibinfo{pages}{2137} (\bibinfo{year}{1990}).

\bibitem[{\citenamefont{Robert}(1991)}]{Robert_1991}
\bibinfo{author}{\bibfnamefont{R.}~\bibnamefont{Robert}}, \bibinfo{journal}{J.
  Stat. Phys.} \textbf{\bibinfo{volume}{65}}, \bibinfo{pages}{531}
  (\bibinfo{year}{1991}).
\bibinfo{author}{\bibfnamefont{R.}~\bibnamefont{Robert}} \bibnamefont{and}
  \bibinfo{author}{\bibfnamefont{J.}~\bibnamefont{Sommeria}},
  \bibinfo{journal}{J. Fluid Mech.} \textbf{\bibinfo{volume}{229}},
  \bibinfo{pages}{291} (\bibinfo{year}{1991}).

\bibitem[{\citenamefont{{Majda} and {Wang}}(2006)}]{Majda_Wang}
\bibinfo{author}{\bibfnamefont{A.}~\bibnamefont{{Majda}}} \bibnamefont{and}
  \bibinfo{author}{\bibfnamefont{X.}~\bibnamefont{{Wang}}},
  \emph{\bibinfo{title}{{Nonlinear Dynamics and Statistical Theories for Basic
  Geophysical Flows}}} (\bibinfo{publisher}{Cambridge University Press},
  \bibinfo{year}{2006}).

\bibitem[{\citenamefont{Bouchet and Sommeria}(2002)}]{Bouchet_Sommeria_2002}
\bibinfo{author}{\bibfnamefont{F.}~\bibnamefont{Bouchet}} \bibnamefont{and}
  \bibinfo{author}{\bibfnamefont{J.}~\bibnamefont{Sommeria}},
  \bibinfo{journal}{J. Fluid Mech.}  (\bibinfo{year}{2002}).

\bibitem[{\citenamefont{Nicolis and Prigogine}(1977)}]{Prigogine}
\bibinfo{author}{\bibfnamefont{G.}~\bibnamefont{Nicolis}} \bibnamefont{and}
  \bibinfo{author}{\bibfnamefont{I.}~\bibnamefont{Prigogine}},
  \emph{\bibinfo{title}{Self-organization in non-equilibrium systems : from
  dissipative structures to order through fluctuations}}
  (\bibinfo{publisher}{Wiley}, \bibinfo{year}{1977}).

\bibitem[{\citenamefont{Robert and Sommeria}(1992)}]{Robert_Sommeria_1992}
\bibinfo{author}{\bibfnamefont{R.}~\bibnamefont{Robert}} \bibnamefont{and}
  \bibinfo{author}{\bibfnamefont{J.}~\bibnamefont{Sommeria}},
  \bibinfo{journal}{Phys. Rev. Lett.} \textbf{\bibinfo{volume}{69}}
  (\bibinfo{year}{1992}).

\bibitem[{\citenamefont{Bouchet and Morita}(2010)}]{Bouchet_Morita_2010}
\bibinfo{author}{\bibfnamefont{F.}~\bibnamefont{Bouchet}} \bibnamefont{and}
  \bibinfo{author}{\bibfnamefont{H.}~\bibnamefont{Morita}},
  \bibinfo{journal}{Physica D} \textbf{\bibinfo{volume}{239}},
  \bibinfo{pages}{948} (\bibinfo{year}{2010}).

\bibitem[{\citenamefont{Morita et~al.}()\citenamefont{Morita, Simonnet, and
  Bouchet}}]{Morita_Simonnet_Bouchet}
\bibinfo{author}{\bibfnamefont{H.}~\bibnamefont{Morita}},
  \bibinfo{author}{\bibfnamefont{E.}~\bibnamefont{Simonnet}}, \bibnamefont{and}
  \bibinfo{author}{\bibfnamefont{F.}~\bibnamefont{Bouchet}}, \bibinfo{note}{in
  preparation}.

\bibitem[{\citenamefont{Okamoto and {Sh\=oji}}(1993)}]{Okamoto_Shoji_1993}
\bibinfo{author}{\bibfnamefont{H.}~\bibnamefont{Okamoto}} \bibnamefont{and}
  \bibinfo{author}{\bibfnamefont{M.}~\bibnamefont{{Sh\=oji}}},
  \bibinfo{journal}{Japan J. Indust. Appl. Math.}
  \textbf{\bibinfo{volume}{10}}, \bibinfo{pages}{191} (\bibinfo{year}{1993}).

\bibitem[{\citenamefont{{Yin} et~al.}(2003)\citenamefont{{Yin}, {Montgomery},
  and {Clercx}}}]{Yin_Montgomery_Clercx_2003}
\bibinfo{author}{\bibfnamefont{Z.}~\bibnamefont{{Yin}}},
  \bibinfo{author}{\bibfnamefont{D.~C.} \bibnamefont{{Montgomery}}},
  \bibnamefont{and} \bibinfo{author}{\bibfnamefont{H.~J.~H.}
  \bibnamefont{{Clercx}}}, \bibinfo{journal}{Physics of Fluids}
  \textbf{\bibinfo{volume}{15}}, \bibinfo{pages}{1937} (\bibinfo{year}{2003}).

\bibitem[{\citenamefont{Bouchet and Simonnet}(2009)}]{Bouchet_Simonnet_2009}
\bibinfo{author}{\bibfnamefont{F.}~\bibnamefont{Bouchet}} \bibnamefont{and}
  \bibinfo{author}{\bibfnamefont{E.}~\bibnamefont{Simonnet}},
  \bibinfo{journal}{Phys. Rev. Lett.} \textbf{\bibinfo{volume}{102}},
  \bibinfo{pages}{094504} (\bibinfo{year}{2009}).

\bibitem[{\citenamefont{Gottlieb and Orszag}(1987)}]{Orszag}
\bibinfo{author}{\bibfnamefont{D.}~\bibnamefont{Gottlieb}} \bibnamefont{and}
  \bibinfo{author}{\bibfnamefont{S.~A.} \bibnamefont{Orszag}},
  \emph{\bibinfo{title}{Numerical analysis of spectral methods: theory and
  applications}} (\bibinfo{publisher}{Society for Industrial Mathematic},
  \bibinfo{year}{1987}).

\bibitem[{\citenamefont{Morita and Kaneko}(2006)}]{Morita_Kaneko_2006}
\bibinfo{author}{\bibfnamefont{H.}~\bibnamefont{Morita}} \bibnamefont{and}
  \bibinfo{author}{\bibfnamefont{K.}~\bibnamefont{Kaneko}},
  \bibinfo{journal}{Phys. Rev. Lett.} \textbf{\bibinfo{volume}{96}},
  \bibinfo{pages}{050602} (\bibinfo{year}{2006}).

\bibitem[{\citenamefont{{Chirikov}}(1979)}]{Chirikov_1979}
\bibinfo{author}{\bibfnamefont{B.~V.} \bibnamefont{{Chirikov}}},
  \bibinfo{journal}{Phys. Rep.} \textbf{\bibinfo{volume}{52}},
  \bibinfo{pages}{263} (\bibinfo{year}{1979}).

\bibitem{small_osc}
\bibinfo{author}{\bibfnamefont{J.~L.} \bibnamefont{{Tennyson}}},
  \bibinfo{author}{\bibfnamefont{J.~D.} \bibnamefont{{Meiss}}},
  \bibnamefont{and} \bibinfo{author}{\bibfnamefont{P.~J.}
  \bibnamefont{{Morrison}}}, \bibinfo{journal}{Physica D}
  \textbf{\bibinfo{volume}{71}}, \bibinfo{pages}{1} (\bibinfo{year}{1994}).
\bibinfo{author}{\bibfnamefont{G.}~\bibnamefont{Manfredi}},
  \bibinfo{journal}{Phys. Rev. Lett.} \textbf{\bibinfo{volume}{79}},
  \bibinfo{pages}{2815} (\bibinfo{year}{1997}).
\bibinfo{author}{\bibfnamefont{C.}~\bibnamefont{Lancellotti}} \bibnamefont{and}
  \bibinfo{author}{\bibfnamefont{J.~J.} \bibnamefont{Dorning}},
  \bibinfo{journal}{Phys. Rev. Lett.} \textbf{\bibinfo{volume}{81}},
  \bibinfo{pages}{5137} (\bibinfo{year}{1998}).
\bibinfo{author}{\bibfnamefont{M.}~\bibnamefont{Brunetti}},
  \bibinfo{author}{\bibfnamefont{F.}~\bibnamefont{Califano}}, \bibnamefont{and}
  \bibinfo{author}{\bibfnamefont{F.}~\bibnamefont{Pegoraro}},
  \bibinfo{journal}{Phys. Rev. E} \textbf{\bibinfo{volume}{62}},
  \bibinfo{pages}{4109} (\bibinfo{year}{2000}).
\bibinfo{author}{\bibfnamefont{J.~R.} \bibnamefont{Danielson}},
  \bibinfo{author}{\bibfnamefont{F.}~\bibnamefont{Anderegg}}, \bibnamefont{and}
  \bibinfo{author}{\bibfnamefont{C.~F.} \bibnamefont{Driscoll}},
  \bibinfo{journal}{Phys. Rev. Lett.} \textbf{\bibinfo{volume}{92}},
  \bibinfo{pages}{245003} (\bibinfo{year}{2004}).
\bibinfo{author}{\bibfnamefont{F.}~\bibnamefont{Valentini}},
  \bibinfo{author}{\bibfnamefont{V.}~\bibnamefont{Carbone}},
  \bibinfo{author}{\bibfnamefont{P.}~\bibnamefont{Veltri}}, \bibnamefont{and}
  \bibinfo{author}{\bibfnamefont{A.}~\bibnamefont{Mangeney}},
  \bibinfo{journal}{Phys. Rev. E} \textbf{\bibinfo{volume}{71}},
  \bibinfo{pages}{017402} (\bibinfo{year}{2005}).
\bibinfo{author}{\bibfnamefont{A.}~\bibnamefont{{Antoniazzi}}}
  \bibnamefont{et~al.}, \bibinfo{journal}{J. Phys. Conf. Ser.}
  \textbf{\bibinfo{volume}{7}}, \bibinfo{pages}{143} (\bibinfo{year}{2005}).

\bibitem[{\citenamefont{Dauxois et~al.}(2010)\citenamefont{Dauxois, Ruffo, and
  Cugliandolo}}]{Dauxois_RUffo_Cugliandolo@LesHouches2008}
\bibinfo{editor}{\bibfnamefont{T.}~\bibnamefont{Dauxois}},
  \bibinfo{editor}{\bibfnamefont{S.}~\bibnamefont{Ruffo}}, \bibnamefont{and}
  \bibinfo{editor}{\bibfnamefont{L.~F.} \bibnamefont{Cugliandolo}}, eds.,
  \emph{\bibinfo{title}{Long-Range Interacting Systems}},
  vol.~\bibinfo{volume}{80} of \emph{\bibinfo{series}{Lecture Notes of the Les
  Houches Summer School}} (\bibinfo{publisher}{Oxford University Press},
  \bibinfo{year}{2010}), \bibinfo{note}{august 2008}.

\end{thebibliography}

\end{document}